\documentclass[11pt]{article}

\usepackage{color}
\usepackage{graphicx}
\usepackage{hyperref}
\usepackage{epsf}
\usepackage{graphicx,epsfig}
\usepackage{cite}
\usepackage{float}
\usepackage{caption}
\usepackage{subcaption}

\usepackage{mathrsfs}
\usepackage{amsmath}
\usepackage{amssymb}
\usepackage{epsfig}
\usepackage{cite}
\usepackage{color,colordvi}
\newcommand{\be}{\begin{equation}}
\newcommand{\ee}{\end{equation}}
\newcommand{\bi}{\begin{itemize}}
\newcommand{\ei}{\end{itemize}}

\definecolor{colGreen}{rgb}{0.1,0.5,0.1}

\newcounter{hran}

\def\MSoverline{\relax\ifmmode\overline{\rm MS}\else{$\overline{\rm MS}${ }}\fi}

\def\vacuum expectation value#1{\left\Big<#1\right\Big>}

\def\ls{\left[}
\def\rs{\right]}

\def\p{\partial}

\def\s{\sigma}

\begin{document}

\thispagestyle{empty}
%
%
\vspace*{-11mm}
%
%

\begin{flushright}
\hfill{ DFPD-2016/TH/06 
} 
\end{flushright}

\vspace*{-1mm}

\begin{center}

\def\thefootnote{\fnsymbol{footnote}}

{\LARGE \bf  Scanning of the  Supersymmetry Breaking Scale \\ and   the Gravitino Mass in Supergravity 
 \\
\vspace{0.25cm}	
}
\vspace{1cm}

{\Large Fotis  Farakos$^{a,b}$,  Alex  Kehagias$^{c}$, Davide Racco$^{d}$ and Antonio  Riotto$^{d}$}
\\[0.5cm]

{\large 
\vspace{.2cm}
{\it  $^{a}$Dipartimento di Fisica ``Galileo Galilei '' \\
Universita di Padova, Via Marzolo 8, 35131 Padova, Italy}\\

\vspace{.2cm}
 {\it  $^{b}$INFN, Sezione di Padova\\
Via Marzolo 8, 35131 Padova, Italy}\\

\vspace{.2cm}
{\it  $^{c}$ Physics Division, National Technical University of Athens, \\15780 Zografou Campus, Athens, Greece}\\

\vspace{.2cm}
{ \it $^{d}$ Department of Theoretical Physics and Center for Astroparticle Physics (CAP)\\ 24 quai E. Ansermet, CH-1211 Geneva 4, Switzerland}\\
}

\vspace{.3cm}


\end{center}

\vspace{1cm}

\hrule \vspace{0.3cm}
 \noindent \textbf{Abstract} \\[0.3cm]
\noindent 
We consider the minimal three-form ${\cal N}=1$ supergravity coupled to nilpotent three-form chiral superfields. 
The supersymmetry breaking is sourced by the three-forms of the chiral multiplets, 
while the value of the gravitino mass is controlled by the three-form of the supergravity multiplet. 
The three-forms  can nucleate membranes which scan both the supersymmetry  breaking scale and  the gravitino mass. The peculiar supergravity feature  that the cosmological constant is the sum of a positive contribution from the supersymmetry breaking scale and 
 a negative contribution from the gravitino mass makes  the cosmological constant jump. This can lead to a phenomenologically allowed  small value of the cosmological constant even though  the supersymmetry breaking scale and the gravitino mass are dynamically large. 
\vspace{0.5cm}  \hrule
\vskip 1cm

\def\thefootnote{\arabic{footnote}}
\setcounter{footnote}{0}


\baselineskip= 15pt

\newpage

\section{Introduction} 
Attempts to solve the cosmological constant problem have triggered many interesting and novel ideas \cite{Weinberg,Witten}. However, it still remains one of the most vexing  and challenging problems.
One possible way to tackle  it  is to make the cosmological constant 
dynamical \cite{Aurilia,Duff,Hawking,BT1,BT2,BP,Feng,DVil}. Usually this is achieved by introducing a three-form field.  Although it has no local dynamics since its field equations demand its four-form field strength to be locally constant,  it contributes to the vacuum energy. Therefore,   it   sources the effective cosmological constant and eventually may   neutralize it.  Moreover, the antisymmetric four-form field strength is coupled to membranes and,  as an electric field can create particle pairs,  the four-form  field may create membranes. 

When the system is coupled  to gravity,  the effective cosmological constant will have two sources: the usual UV contribution   and the contribution  from the four-form energy density, which is a constant by its field equations. 
The effective cosmological constant is then the algebraic sum of these two sources. Its value can be lowered by   the 
spontaneous creation of closed membranes. In other words, the nucleation of membranes in the constant four-form backgrounds tends to reduce the effective  cosmological constant and hopefully drive the latter to its observed value. This mechanism for the scanning of the  cosmological constant through membrane nucleation has been proposed by Brown and Teitelboim (BT) \cite{BT1,BT2}. It is  is similar to the Schwinger effect in which electron-positron pairs are spontaneously produced in a constant electric field in such a way as to reduce the value of the latter.  The membrane nucleation will stop as soon as the cosmological constant is very close to zero such that further nucleation will lead to an anti-de Sitter bubble, a process which is very suppressed \cite{BT2}. The final value of the cosmological constant is controlled by the membrane tension and its charge, and, for sufficiently small values of the latter, we can have a cosmological constant as small as the observed one. Therefore, membrane nucleation provides a dynamical mechanism for reducing the value of the cosmological constant \cite{BT1,BT2,BP}. 

Unfortunately, the mechanism, although appealing, has two serious drawbacks \cite{BT1,BT2,BP}. The first one is that the value of the membrane charge needed  to end up with  the observed value of the cosmological constant should be  so tiny that it is beyond any reasonable justification  of microscopic origin. The second problem is the empty universe.  Membrane nucleation occurs during  a prolonged de Sitter phase, thus leading to an empty universe,  very different from the one we observe. 

The above aforementioned problems are rather discouraging for a true solution to the cosmological constant problem. However, an elegant solution was provided  as a variant of the BT mechanism  by Bousso and Polchinski (BP)\cite{BP}, which involves many four-forms. The latter are coupled to membranes with (incommensurate)  charges which form a grid leading to an appropriately dense discretum,  where the observed cosmological constant 
is accommodated. In this case, the observed cosmological constant is proportional to the volume of the fundamental cell of the grid and inversely proportional to an appropriate power of  the overlined cosmological constant (the one without the contribution of the four-forms). Therefore, the smallness of the cosmological constant is attributed not to the   tininess of a membrane  charge, but rather to the smallness of the
volume of the fundamental shell in  the charge grid. In this case for example, an overlined cosmological constant  at the Planck scale $M_p=1/\sqrt{8\pi G}$ can be dynamically driven to zero for a  hundred of membranes with charges of the order of $2\times 10^{-2}$ in Planck units. 

In addition, in this multi four-form BP setup, the empty universe is easily solved by a slow roll inflaton field \cite{BP}.  Since now the last transition does not necessarily  happen between bubbles with small cosmological constants, in the penultimate step the cosmological constant can be large. In this case, quantum displacement of the inflaton field may dominate over  the one induced by the classical motion, thus  giving the necessary kick off to the inflaton to be away from its minimum. Then, during the last step, the inflaton  slowly rolls towards its minimum where it  starts oscillating, ultimately reheating  the universe.  

The goal of this paper is to  point out an interesting interplay between the scanning of the cosmological constant discussed above and the ones of the supersymmetry breaking scale and of the gravitino mass in supergravity. 
In  global supersymmetry, four-forms  reside in the imaginary  part  of the higher component of a chiral superfield. Therefore, when they get a vacuum expectation value, they contribute both to the cosmological constant as well as to the supersymmetry breaking scale. In such a  case, membrane nucleation will not only reduce the value of the cosmological constant, but will also reduce the supersymmetry breaking scale, leading to an almost zero cosmological constant for  a supersymmetric universe. 

\begin{figure}[h]
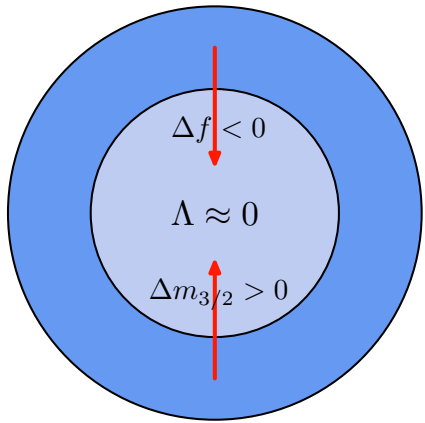

    \centering
    $\hbox{ \convertMPtoPDF{membranes.1}{1.1}{1.1} }$
    \caption{ A schematic view of the mechanism described in the paper. }
    \label{fig:awesome_image1}
\end{figure}
\noindent
In supergravity things are different. There are two competitive contributions to the cosmological constant, a positive one 
proportional to the  square of the  supersymmetry breaking scale $f$, which increases its value, and a negative one from the square of the 
gravitino mass which works towards reducing it ,
 \begin{equation}
 \Lambda=f^2-3 \, m_{3/2}^2 \,.
 \end{equation}
Hence, the cosmological constant can be reduced to zero by decreasing the values of  supersymmetry breaking scale $f$ and increasing the gravitino mass.  The  membrane nucleation process may include not only the scanning of   the supersymmetry breaking scale,  but also  the scanning of the gravitino mass, in such a way that they approach each other during the nucleation process. 
The scanning of both $f$ and $m_{3/2}$ 
induces a scanning of the 
 cosmological constant 
 and can be such that they remain high although $\Lambda$ is driven to zero, as depicted in Fig.~\ref{fig:awesome_image1}. 
Indeed, the  nucleations stop at the point where the cosmological constant is close to zero and a further jump into anti-de Sitter is forbidden, and, at the same time, the supersymmetry breaking scale and the gravitino mass are large. 
 This is the key observation of this paper.

The mechanism works by having both  the supersymmetry breaking scale $f$ and the gravitino mass $m_{3/2}$ make jumps\footnote{Supersymmetry breaking scale scanning has been considered also in Ref.~\cite{BGM}, albeit due  QCD instanton effects linked to the relaxation mechanism \cite{relaxion}.}. 
The gravitino mass may be rendered a scanning variable by relating it to the vacuum expectation value of a four-form, which replaces one of the two auxiliary scalars of the minimal ${\cal N}=1$ supergravity. This three-form, although it has no local dynamics (it replaces an auxiliary field), can scan a range of possible values as it couples to a (supersymmetric) membrane. This leads to a corresponding scanning of the gravitino mass which tends then to neutralize the cosmological constant emerging from the supersymmetry breaking.

The supersymmetry breaking scale is rendered a scanning variable in the following way. 
First, supersymmetry is broken by a three-form chiral superfield $\Phi$. 
Once  we break   supersymmetry,  
we can impose a nilpotency  condition\footnote{The use of the nilpotency condition  has been employed only  for simplicity reasons. We could  easily consider unconstrained superfields at the cost of unnecessary complications. Hence, for simplicity, we have decided to decouple the scalar partners of the three-form multiplets.} on the chiral superfield $\Phi$  \cite{Rocek,Lindstrom,Casalbuoni,Seiberg}
\begin{equation}
 \Phi^2=0,  \label{nil0}
\end{equation}
 which removes the scalar lowest component from the spectrum. 
 Secondly, 
 we introduce additional  three-forms residing 
 in chiral multiplets $Y^i$ and we 
 consistently remove their scalar partner by imposing the constraints \cite{Brignole:1997pe,Seiberg,Dall'Agata:2015zla,Vercnocke:2016fbt} 
 \begin{equation}
  \Phi Y^i=0.  \label{nil10}
 \end{equation}
These constraints remove the lowest scalar components 
of the chiral multiplets. 
The superfields $Y^i$ also contribute to the supersymmetry breaking. 
The scanning of both the supersymmetry breaking scale as well as the gravitino mass 
make the mechanism described above work towards an almost vanishing cosmological constant. 

All in all,  the minimal  content of the theory for the mechanism to work  is the following:
\begin{itemize}
\item The three-form ${\cal N}=1$ supergravity multiplet. This is basically the usual supergravity multiplet in which one of two auxiliary scalars of the off-shell multiplet is traded for the dual of a four-form field strength.
\item The chiral superfield $\Phi$, the highest component of which 
is a four-form field strength and gets a {\rm vacuum expectation value}. 
We decouple the scalar of $\Phi$ by the constraint (\ref{nil0}). The fermion of $\Phi$ will contribute to the goldstino. 
\item  Three-form multiplets $Y^i$. These are chiral multiplets in which one of the scalars in their highest components is traded for a four-form field strength and  gets a {\rm vacuum expectation value}. 
We  impose the condition (\ref{nil10}) on $Y^i$ for simplicity. The fermion of the $Y^i$ will contribute to the goldstino. 
\item The inflaton multiplet ${\cal A}$. This is a chiral multiplet, whose real part of its lower component is the inflaton field, needed to reheat the universe. 
\end{itemize}
The structure of the paper is as follows: In section 2 we present the formulation of the minimal ${\cal N}=1$ supergravity in terms of a three-form auxiliary field. In section 3 we discuss the three-form chiral multiplets and their couplings to supergravity and  membranes. In section 4 we present how the supersymmetry breaking, the gravitino mass and the cosmological constant are scanned by membrane nucleation in the case of multi four-forms, and finally we conclude in section 5. 
We use $M_p =1$ units and the conventions of \cite{Wess:1992cp}.

\section{Three-form chiral superfields and  three-form supergravity}

In this work we will utilize three-form multiplets for the supersymmetry breaking sector.  
The breaking works much the same way as for the  Polonyi multiplet,  
but here the supersymmetry breaking scale will be related to the vacuum expectation value of the field strength of the three-forms.

The three-form multiplet \cite{Gates:1980ay,Gates:1983nr,Binetruy:1996xw}  resides in  a chiral superfield 
\be
\overline {\cal D}_{\dot \alpha} \Phi = 0, 
\ee
with component definitions 
\be
\label{3m}
\begin{split}
\Phi | =& A,  
\\
{\cal D}_\alpha \Phi | =& \sqrt{2} \chi_\alpha,
\\
{\cal D}^2 \Phi| +  \overline {\cal D}^2 \overline \Phi |  = & - 8 F_1,  
\\
{\cal D}^2 \Phi| - \overline {\cal D}^2\overline  \Phi|  
=&    \frac{32 i}{3} \epsilon^{mnpq} \p_m C_{npq} 
+ 2 \sqrt{2} i\overline  \psi_m\overline  \s^m \chi 
 - 2 \sqrt{2} i \psi_m \s^m \overline  \chi 
\\
& 
-4 \left( \overline  M + \overline \psi_m \overline \s^{mn}\overline \psi_n \right) A 
+  4 \left( M + \psi_m \s^{mn} \psi_n \right) \overline  A. 
\end{split}
\ee
Here $F_1$ is a real auxiliary scalar, $C_{mnp}$ is a real three-form and $M$ is the supergravity auxiliary field. 
In total, the chiral multiplet contains the component fields
\be
(A,~~\chi_\alpha ,  ~~F_1 , ~~ C_{mnp}). 
\ee 
One can define a  three-form real superfield ${\cal C}_{ABC}$, 
whose lowest components vanish in the appropriate Wess-Zumino gauge \cite{Gates:1980ay,Binetruy:1996xw}, 
except the component 
\be
{\cal C}_{abc} | = C_{abc} . 
\ee 
It is important to recall that  the supersymmetry transformation of  $\chi_\alpha$  is 
\be
\delta \chi_\alpha = - \sqrt 2 F \zeta_\alpha 
- i \sqrt 2 \s^m_{\alpha \dot \beta} \overline \zeta^{\dot \beta} \left( \p_m A - \frac{1}{\sqrt 2}\psi_{m}^{\ \beta} \chi_\beta   \right),   \label{cc}
\ee
where 
\be
\begin{split}
\label{F1}
F = F_1 &+ i \left( -\frac43 \epsilon^{mnpq} \p_m C_{npq} 
-\frac{1}{2 \sqrt{2}} \overline \psi_m\overline \s^m \chi 
+ \frac{1}{2 \sqrt{2}}\psi_m \s^m\overline \chi   \right) 
\\
&- \frac{i}{2} \Big{[} \left( \overline M + \overline  \psi_m \overline  \s^{mn} \overline  \psi_n \right) A 
- \left( M + \psi_m \s^{mn} \psi_n \right)\overline A \Big{]}.
\end{split}
\ee
From the transformation (\ref{cc})  we see that when 
\be
\label{vacuum expectation value1}
\langle  \epsilon^{mnpq} \p_m C_{npq}  \rangle \ne 0 
\ee 
the goldstino is  $\chi_\alpha$ since 
\be
\delta \chi_\alpha =  \frac{4 \sqrt 2 i}{3}  \langle  \epsilon^{mnpq} \p_m C_{npq}  \rangle \zeta_\alpha  + \cdots 
\ee 
In addition, once supersymmetry is  broken, 
it is possible to consistently remove the sgoldstino (here the scalar $A$) from the spectrum, 
which simplifies the model. 
We  impose a nilpotency condition on the chiral three-form superfield 
\be
\label{c1}
\Phi^2 = 0 . 
\ee
Taking into account that 
\be
\Phi = A + \sqrt 2 \Theta \chi + \Theta^2 F, 
\ee
where $F$ is given in \eqref{F1}, the constraint \eqref{c1} is solved 
by eliminating $A$ in terms of the other component fields of the theory 
\be
\label{Anil}
A = \frac{\chi^\alpha \chi_\alpha}{2 F_1 - \frac{8 i}{3} \epsilon^{mnpq} \p_m C_{npq} 
+ \frac{i}{2 \sqrt 2} \psi^a \s_a \overline \chi 
-(M+ \psi_c \s^{cd} \psi_d) \overline \chi^2)/4(F_1 + \frac{4i}{3} \epsilon^{mnpq} \p_m C_{npq} ) } 
\ee
which is non-singular as long as \eqref{vacuum expectation value1} holds.

Let us also rapidly review the structure of the minimal three-form ${\cal  N}=1$ supergravity \cite{Ovrut:1997ur}. 
This  theory originates from the old-minimal ${\cal  N}=1$ supergravity by replacing one of the two real scalar auxiliary fields of the graviton multiplet by a three-form. 
In this case, the free theory is anti-de Sitter supergravity where the mass of the gravitino is actually the vacuum expectation value  of the four-form field strength and the cosmological constant is negative and proportional to the square of the gravitino mass \cite{Ovrut:1997ur}.

The three-form supergravity  is constructed by starting with an additional three-form superfield $\Phi'$, 
with component fields definitions as in \eqref{3m}, 
but imposing on it the constraint 
\be
\label{f'}
\Phi' = - \frac43 .  
\ee 
Let us call the three-form field of the $\Phi'$ multiplet $H_{mnp}$. 
The constraint \eqref{f'} has two effects: 
first it removes all the component fields of $\Phi'$, 
except the real three-form, $H_{mnp}$ and
secondly it implies that for the complex auxiliary $M=(M_1+i M_2)$ of the supergravity multiplet 
we have
\be
\label{3sugra}
M_2 = \epsilon^{mnpq} \p_m H_{npq} + \frac{i}{2} \left(  \psi_a \s^{ab} \psi_b -\overline  \psi_a \s^{ab}\overline \psi_b  \right).    
\ee
The supergravity multiplet now contains the component fields 
\be
(e^a_m,~~\psi_m^\alpha ,  ~~M_1 , ~~b_m, ~~ H_{mnp}), 
\ee
which again form a minimal set of degrees of freedom. 
Note that one can again define a  three-form real superfield ${\cal H}_{ABC}$, 
whose  lowest components vanish in the Wess-Zumino gauge, except  of 
\be
{\cal H}_{abc} | =  H_{abc} . 
\ee

\section{Coupling to membranes}

In this section we couple a single three-form nilpotent chiral superfield $\Phi$ to the three-form supergravity. 
In addition,  the three-forms of the supergravity multiplet and of the chiral superfield $\Phi$ 
are coupled to fundamental membranes \cite{Bergshoeff:1987cm,Ovrut:1997ur,DF}. 
The action describing this model is  
\be
S = S_{SG} + S_{B} .  
\ee
For the supergravity and three-form multiplet sector we have  
\be
\label{L1}
{\cal L}_{SG} = -\frac18 \int {\rm d}^2 \Theta \, 2 {\cal E} \, \ls  \left(\overline {\cal D}^2 -8 {\cal R} \right)  \Omega(\Phi, \overline \Phi) \rs + {\rm c.c.}, 
\ee 
where $\Omega$ is a hermitian function and there is no superpotential. 
For the  membrane sector we have  
\be
\label{LB1}
\begin{split}
S_{B} = \ & T_3 \int {\rm d}^3 \xi \sqrt{-\gamma} \left( -\frac12 \gamma^{ij} \Pi_i^a \Pi_j^b \eta_{ab} 
+\frac12 
+  \epsilon^{ijk} \Pi_i^A \Pi_j^B \Pi_k^C {\cal H}_{CBA} \right) 
\\
+&  \widetilde{T}_3 \int {\rm d}^3 \zeta \sqrt{- \widetilde \gamma} \left( -\frac12 \widetilde \gamma^{ij} T_i^a T_j^b \eta_{ab} 
+\frac12 
+ \frac{4}{\sqrt{3}}   \epsilon^{ijk} T_i^A T_j^B T_k^C {\cal C}_{CBA} \right),
\end{split}
\ee
where 
\be
\Pi_i^A = \frac{\p Z^M}{\p \xi^i} \, E_{M}^{A}  \ , \ T_i^A = \frac{\p \widetilde Z^M}{\p \zeta^i} \, E_{M}^{A} 
\ee
and $Z^M$ and $\widetilde Z^M$ are the coordinates of the membranes, with $i=0,1,2$. 
The membrane action also contains the coupling of the supergravity 
three-form $H_{mnp}$ and the matter three-form $C_{mnp}$ to the membranes. 
Further discussion on three-forms in supersymmetry and supergravity can 
be found in \cite{Bielleman:2015ina,Groh:2012tf,Bandos:2012gz,Bandos:2011fw,Farrar:1997fn,Dudas:2014pva,Dvali:1999pk,Dvali:2005an}. 
Here we have to pause and point out that the membrane coupled to the supergravity three-form has a $\kappa$-symmetry, 
while the one coupled to the matter three-form does not. 
Therefore, on the membrane coupled to $H_{mnp}$ half supersymmetry is preserved, 
while on the membrane coupled to $C_{mnp}$, no supersymmetry is preserved. 
This should not be confused with the bulk supersymmetry away from the branes which is manifestly preserved.

The bosonic sector after we integrate out the supergravity auxiliary field $b_m$ reads 
\be
\begin{split}
\label{L1}
S = & \int \! {\rm d}^4x \,e \Big{[} \frac{\Omega}{6} R 
- \Omega_{A \overline  A} \p A \p \overline A 
-\frac{1}{4 \Omega} (\Omega_A \p_m A - \Omega_{\overline  A} \p_m \overline  A  )(\Omega_A \p_m A - \Omega_{\overline  A} \p_m \overline  A ) 
\\
&\ \ \ \ \ \ 
\ \ \ \ \ 
+ \frac{\Omega}{9} M \overline M 
+ \Omega_{A \overline  A} F \overline F 
- \frac13 \Omega_A M F 
- \frac13 \Omega_{\overline  A} \overline  M\, \overline F \Big{]} 
\\
& + T_3 \int {\rm d}^3 \xi \sqrt{-\gamma} \left( -\frac12 \gamma^{ij} \p_i X^m \p_j X^n g_{mn} 
+\frac12 
+  \epsilon^{ijk} \p_i X^m \p_j X^n \p_k X^p  H_{mnp} \right) 
\\
& + \widetilde T_3 \int {\rm d}^3 \zeta \sqrt{- \widetilde \gamma} \left( -\frac12 \widetilde \gamma^{ij} \p_i \widetilde X^m \p_j \widetilde X^n g_{mn} 
+\frac12 
+ \frac{4}{\sqrt{3}}  \epsilon^{ijk} \p_i \widetilde X^m \p_j \widetilde X^n \p_k \widetilde X^p  C_{mnp} \right). 
\end{split}
\ee
Here $F$ and $M$ contain the three-forms in their imaginary parts, 
and we have not dropped the composite scalar $A$ given by \eqref{Anil} for the moment. 
Note that $F$ will contain both the three-form of its own multiplet but also the supergravity one. 
For the $\Omega$ function we choose the following expression  
\be
\label{omega1}
\Omega = -3 +  \Phi \overline \Phi  
\ee
and  $\Phi$ satisfies the nilpotency condition and we disregarded linear pieces in $\Phi$ as they do not alter our findings.  
As we will see in this setup Eq. \eqref{vacuum expectation value1} always holds, 
therefore the Goldstone mode will always be $\chi_\alpha$. 
Being the goldstino, this
means that   $\chi_\alpha$ is a gauge degree of freedom
and  we can go to a gauge where 
\be
\label{simp}
 \chi_\alpha = 0 . 
\ee 
In the gauge \eqref{simp} the complete supergravity Lagrangian is 
\be
\begin{split}
\label{Lsugra1}
e^{-1} {\cal L}_{SG} = & -\frac12 R 
+ \epsilon^{klmn} \overline  \psi_k\overline  \s_l {\cal D}_m \psi_n 
-\frac{1}{3} (M_1)^2  
+ (F_1)^2 
\\
& 
- \frac{1}{3} (\epsilon^{mnpq} \p_m H_{npq})^2 
- \frac{i}{3} (\epsilon^{mnpq} \p_m H_{npq}) (\psi_c \s^{cd} \psi_d-\overline \psi_c \overline \s^{cd} \overline \psi_d) 
\\
&
+\frac{16}{9}   (\epsilon^{mnpq} \p_m C_{npq})^2 
+ \frac{1}{12} (\psi_c \s^{cd} \psi_d\overline  \psi_c\overline  \s^{cd} \overline \psi_d)^2 . 
\end{split}
\ee
Now we can integrate out the scalar auxiliary fields, 
which give 
\be
F_1 = 0 = M_1 
\ee
as anticipated, 
since the supersymmetry breaking and the gravitino mass originate from the three-forms. 
We redefine the matter three-form as 
\be
\label{redef33}
C_{mnp} = \frac{\sqrt{3}}{4 } \ \Gamma_{mnp} 
\ee
and we define 
four-form field strengths $F_{mnpq}$ and $G_{mnpq}$ as 
\begin{eqnarray}
G_{mnpq}&=&\partial_m H_{npq}-\partial_q H_{mnp}+\partial_p H_{qmn}-\partial_n H_{pqm},\\
F_{mnpq}&=&\partial_m \Gamma_{npq}-\partial_q \Gamma_{mnp}+\partial_p \Gamma_{qmn}-\partial_n \Gamma_{pqm} . 
\end{eqnarray} 
After an appropriate chiral rotation of the gravitino, we may write the Lagrangian (\ref{Lsugra1}) as 
\be
\label{sugra333}
\begin{split}
e^{-1} {\cal L}_{SG} = 
& -\frac12 R 
+ \epsilon^{klmn} \overline  \psi_k \overline  \s_l {\cal D}_m \psi_n 
- \frac{1}{2}  F_{mnpq} F^{mnpq} +\frac{1}{2}G_{mnpq}  G^{mnpq}  
\\
&
+\frac{1}{12}(\epsilon^{mnpq} G_{mnpq}) (\psi_c \s^{cd} \psi_d+\overline \psi_c \overline  \s^{cd} \overline \psi_d) 
- \frac{1}{12} (\psi_c \s^{cd} \psi_d + \overline  \psi_c \overline  \s^{cd} \overline  \psi_d)^2 . 
\end{split}
\ee
From  Eq. \eqref{sugra333},  
comparing with the standard supergravity breaking pattern  \cite{Wess:1992cp}, 
we can easily identify the gravitino mass 
\be
m_{3/2} = \frac13  \langle \epsilon^{mnpq} \p_m H_{npq} \rangle = \frac{1}{12} \langle \epsilon^{mnpq} G_{mnpq} \rangle,
\ee
 the supersymmetry breaking scale 
\be
f^2 = \frac{1}{3}\langle (\epsilon^{mnpq} \p_m \Gamma_{npq} )^2\rangle=  \frac{1}{48} \langle (\epsilon^{mnpq} F_{mnpq})^2 \rangle,
\ee 
and the cosmological constant 
\begin{eqnarray}
\Lambda=\frac{1}{2}\langle G^{mnpq}G_{mnpq} - F^{mnpq}F_{mnpq}\rangle=
f^2-3m_{3/2}^2.
\end{eqnarray} 
The equations of motion for the three-forms $A_{mnp}$ and $\Gamma_{mnp}$ are 
\begin{eqnarray}
&&\nabla_m G^{mnpq}=\epsilon^{mnpq}\partial_m G=\frac{T_3}{4} \int d^3\xi~\delta^{(4)}\Big{(}X-X(\xi)\Big{)}
 \frac{\partial X^n}{\partial \xi^a}
 \frac{\partial X^p}{\partial \xi^b}
 \frac{\partial X^q}{\partial \xi^c} \epsilon^{abc}, \label{eqF1}\\
 &&\nabla_m F^{mnpq}=\epsilon^{mnpq}\partial_m F=-\frac{\widetilde T_3}{4} \int d^3\zeta~\delta^{(4)}\Big{(}X-\widetilde X(\zeta)\Big{)}
 \frac{\partial X^n}{\partial \zeta^a}
 \frac{\partial X^p}{\partial \zeta^b}
 \frac{\partial X^q}{\partial \zeta^c} \epsilon^{abc} \label{eqG1} , 
\end{eqnarray}
where we have expressed the four-forms as 
\begin{eqnarray}
F_{mnpq}=F \, \epsilon_{mnpq},~~~G_{mnpq}=G \,\epsilon_{mnpq}.
\end{eqnarray}
Integrating \eqref{eqF1} and \eqref{eqG1}, we find that the four-forms are discontinuous  across the membranes with finite jumps
\begin{eqnarray}
 \Delta G=-\frac{T_3}{4}\epsilon_{\rm G}, ~~~~\Delta F=\frac{ \widetilde T_3}{4}\epsilon_{\rm F},
 \end{eqnarray} 
 where $\epsilon_{\rm G,F}=\pm 1$ \cite{BT2}.
These conditions are solved by 
\begin{eqnarray}
&&G=G_0+\left| \frac{T_3}{4} \right|\epsilon_{\rm G}  \, n =G_0+\frac{T_3}{4}\epsilon_{\rm G} \, n,\\ &&F=F_0-\left|\frac{\widetilde T_3}{4}\right| \epsilon_{\rm F}   \, \widetilde n =F_0-\frac{\widetilde T_3}{4}\epsilon_{\rm F} \, \widetilde n, 
\end{eqnarray}
where $n$ is the number of nucleation steps, $F_0$ and $G_0$ are the initial values of the four forms, 
$\epsilon_{\rm G}=\epsilon_{\rm F}=-1$ are selected by the dynamics  such that the cosmological constant is reduced at every step \cite{BT2}, 
and $T_3, \widetilde T_3>0$. 
This  will lead to a scanning of the gravitino mass, the supersymmetry breaking scale and the cosmological constant, which are now 
\begin{eqnarray}
&&m_{3/2}^{\phantom{0}}=m_0+\frac{1}{2} T_3 \, n,\\
&&f=f_0-\frac{\sqrt{3}}{2} \widetilde T_3 \, \widetilde n,\\
&&\Lambda= \left( f_0-\frac{\sqrt{3}}{2} \widetilde T_3 \, \widetilde n \right)^2-3 \, \left( m_0+\frac{1}{2} T_3 \, n\right)^2,  
\end{eqnarray}
where we have taken $F_0<0,G_0<0$ and 
\begin{equation}
m_0=2 \Big|G_0 \Big|, ~~~f_0=2\sqrt{3}\Big|F_0 \Big|.  
\end{equation}
From  here we see that a mechanism similar to \cite{BT1,BT2,BP}  is already at work, 
which will reduce the value of the cosmological constant through subsequent membrane nucleations 
since we have a two-dimensional grid: $(n,\widetilde n)$. 
What is novel in this scenario is that not only the cosmological constant is scanned, 
but also the gravitino mass\footnote{ 
Notice that, had  we used the old-minimal supergravity \cite{Wess:1992cp} instead of the three-form supergravity, 
the gravitino mass would not scan. 
Therefore, the cosmological constant would be given by 
$$
\Lambda_\text{old-minimal} = \left( f_0-\frac{\sqrt{3}}{2} \widetilde T_3 \, \widetilde n \right)^2-3 \, m_0^2 , 
$$
where $m_0$ is the gravitino mass, a mass parameter one has to put in the theory with 
the use of a constant  superpotential $W=  m_0$.}.

\section{Multiple membranes and multiple three-forms}

In this section we couple a set of  three-form chiral superfields $Y^I$ 
\be
\overline {\cal D}_{\dot \alpha} Y^I = 0, 
\ee 
where $I=1,\ldots, N$ (and $N$ is the total number of chiral multiplets) 
to the minimal ${\cal N}=1$ three-form supergravity. 
The component definitions are 
\be
\label{3mY}
\begin{split}
Y^I | =& A^I,  
\\
{\cal D}_\alpha Y^I | =& \sqrt{2} \chi^I_\alpha,
\\
{\cal D}^2 Y^I | +  \overline {\cal D}^2 \overline Y^I |  = & - 8 F^I_1,  
\\
{\cal D}^2 Y^I | -  \overline {\cal D}^2 \overline Y^I |  
=&    \frac{32 i}{3} \epsilon^{mnpq} \p_m C^I_{npq} 
+ 2 \sqrt{2} i \overline  \psi_m \overline \s^m \chi^I 
 - 2 \sqrt{2} i \psi_m \s^m \overline  \chi^I 
\\
& 
-4 \left( \overline M + \overline  \psi_m\overline  \s^{mn}\overline  \psi_n \right) A^I 
+  4 \left( M + \psi_m \s^{mn} \psi_n \right) \overline A^I . 
\end{split}
\ee 
The real scalars $F_1^I$ are auxiliary fields and $C_{mnp}^I$ are the real  three-forms of the chiral superfields. 
The three-form fields of the supergravity multiplet and of the chiral superfields $Y^I$ 
will be  coupled to  multiple fundamental membranes \cite{Bergshoeff:1987cm,DF}.

As we will see the supersymmetry breaking in this model comes from  
\be
\label{vacuum expectation value11}
\langle  \epsilon^{mnpq} \p_m C^I_{npq}  \rangle \ne 0 \ , \ \langle F^I_1 \rangle = 0. 
\ee 
Therefore  the goldstino will be a linear combination of $\chi^I_\alpha$. 
In such a setup, 
it is possible to consistently remove all the scalar fields $A^I$ from the theory. 
This will dramatically  simplify the model.  
To do this we break the collective treatment of three-form superfields, by defining 
\be
Y^I = \left( \Phi , \, Y^i \right) . 
\ee 
We impose a  nilpotency condition on the  three-form superfield $\Phi$ of the form 
\be
\label{c11}
\Phi^2 = 0 
\ee 
which removes the scalar $A$ from the spectrum. 
For the other three-form superfields we impose the constraints 
\be
\label{c12}
\Phi \,  Y^i = 0 
\ee
which remove the rest of the scalars $A^i$  from the spectrum. 
To wrap it up, 
by imposing the constraints \eqref{c11} and \eqref{c12}, 
all the scalars $A^I$ are eliminated, 
and the $Y^I$ sector of the  theory will contain only the propagating fermions $\chi^I_\alpha$, 
the real auxiliary fields $F^I_1$ and the real three-forms $C_{mnp}^I$. 
Interestingly, such a system was discussed earlier for standard chiral superfields  in Ref. \cite{Dall'Agata:2015zla}.

For the supergravity and three-form multiplets sectors we will  have  that
\be
\label{L1}
{\cal L}_{SG} = -\frac18 \int {\rm d}^2 \Theta \, 2 {\cal E} \, \ls  \left( \overline  {\cal D}^2 -8 {\cal R} \right)  \Omega(Y^I , \overline Y^J) \rs + {\rm c.c.},
\ee 
where  $\Omega$ is the hermitian function 
\be
\label{omega11}
\Omega = -3 + \delta_{I  J}  Y^I \overline Y^J.   
\ee 
The membrane action will  contain,  as before,  
the coupling of the supergravity three-form $H_{mnp}$ to a single membrane, 
while the coupling of the other three-forms with multiple membranes will be given by 
a generalization of Eq.(\ref{LB1}). 
Each three-form $C_{mnp}^I$ is coupled to a  distinct membrane with tension $T_3^I$. 
Ignoring all spin-$\frac12$ fields,  the supergravity Lagrangian is 
\be
\begin{split}
\label{Lsugra111}
e^{-1} {\cal L}_{SG} = & -\frac12 R 
+ \epsilon^{klmn} \overline  \psi_k \overline \s_l {\cal D}_m \psi_n 
-\frac{1}{3} (M_1)^2  
+ \delta_{I J} (F_1)^I (F_1)^J 
\\
& 
- \frac{1}{3} (\epsilon^{mnpq} \p_m H_{npq})^2 
- \frac{1}{3} (\epsilon^{mnpq} \p_m H_{npq}) (\psi_c \s^{cd} \psi_d + \overline \psi_c \overline \s^{cd} \overline \psi_d) 
\\
&
+\frac{16}{9} \delta_{I J}  \epsilon^{mnpq} \p_m C_{npq}^I \epsilon^{mnpq} \p_m C_{npq}^J  
- \frac{1}{12} (\psi_c \s^{cd} \psi_d + \overline \psi_c \overline  \s^{cd}\overline \psi_d)^2 . 
\end{split}
\ee
In \eqref{Lsugra111} we have performed a chiral rotation in the gravitino to bring the couplings to canonical form. 
Now we can integrate out the scalar auxiliary fields, 
which give 
\be
F^I_1 = 0 = M_1 
\ee
as anticipated. 
We redefine the matter three-forms as 
\be
\label{redef332}
C^I_{mnp} = \frac{\sqrt{3}}{4 } \ \Gamma^I_{mnp} 
\ee
and we define 
four-form field strengths $F^I_{mnpq}, ~G_{mnpq}$ as 
\begin{eqnarray}
G_{mnpq}&=&\partial_m H_{npq}-\partial_q H_{mnp}+\partial_pH_{qmn}-\partial_n H_{pqm},\\
F^I_{mnpq}&=&\partial_m \Gamma^I_{npq}-\partial_q \Gamma^I_{mnp}+\partial_p \Gamma^I_{qmn}-\partial_n \Gamma^I_{pqm}. 
\end{eqnarray} 
Proceeding as previously, 
once we express the four-forms as 
\begin{eqnarray}
F^I_{mnpq}=F^I \, \epsilon_{mnpq},~~~G_{mnpq}=G \,\epsilon_{mnpq} 
\end{eqnarray}
and we integrate the $H_{mnp} $ and $\Gamma^I_{mnp} $ equations of motion, 
we find that the four-forms are discontinuous  across the membranes with finite jumps 
\begin{eqnarray}
 \Delta G=\frac{T_3}{4}, ~~~~\Delta F^I =-\frac{T^I_3}{4}.
 \end{eqnarray} 
These conditions are solved by 
\begin{eqnarray}
&&G=G_0-\frac{T_3}{4}\, n,\\ &&F^I=F^I_0+\frac{T^I_3}{4}\, n^I, 
\end{eqnarray}
where $F^I_0<0$ and $G_0<0$ are the initial values of the four forms and $T_3, T^I_3>0$. 
By using the definitions 
\begin{eqnarray}
m_0=2\Big| G_0 \Big|, ~~~f^I_0=2\sqrt{3} \Big| F^I_0\Big|,  
\end{eqnarray}
we find 
\be
\begin{split}
m_{3/2} & = m_0 + \frac12 T_3 \, n, 
\\
f^I & = f^I_0 - \frac{\sqrt{3}}{2}  T^I_3 \, n^I, 
\end{split}
\ee
and the cosmological constant 
\be
\label{cosmo1}
\Lambda = \sum_{I=1}^N \left( f^I_0 - \frac{\sqrt{3}}{2}  T^I_3 \, n^I   \right)^2 
- 3 \left(  m_0 + \frac12 T_3 \, n  \right)^2. 
\ee 
From \eqref{cosmo1} we see that, after a series of successive membrane nucleations, the cosmological constant is neutralized, 
exactly as it happens in the BP framework, with a grid of steps 
\be
n_A = (n, n_I ) . 
\ee 
For simplicity we will assume, also in order to keep contact with the notation of \cite{BP}, that the four-forms charges are quantized. This is not generally necessary, but it is the case if the four-forms originate from a higher dimensional theory like string- or M-theory. 
Then, we may express \eqref{cosmo1} as 
\begin{eqnarray}
\label{cosmo2}
\Lambda = \sum_{I=1}^N \left(   n^I q^I   \right)^2 
-  \left(   n^0q^0  \right)^2,  
\end{eqnarray}
where 
\begin{eqnarray}
q^I=\frac{\sqrt{3}}{2}  T^I_3, ~~~q_0=\frac{\sqrt{3}}{2}  T_3, 
\end{eqnarray}
and the initial values of $(n^0,n^I)$ are 
\begin{eqnarray}
n^0_{\rm int}=\frac{2 m_0}{ T_3},~~~n^I_{\rm int}=\frac{2f_0}{\sqrt{3} T^I_3}.
\end{eqnarray}
In this case, we start with a large initial $n^A_{\rm int}=(n^0_{\rm int},n^I_{\rm int})$ and we scan the grid towards lower values of $n^A$. Let us note that we may write the cosmological constant \eqref{cosmo2} 
as 
\begin{eqnarray}
\Lambda=\sum_{A=0}^N Q^A \eta_{AB} Q^B ,~~~~Q^A=q^A n^A, ~~~\eta_{AB}={\rm diag}\left(-1,1,\ldots1\right). \label{cosmo3}
\end{eqnarray}
Therefore $\Lambda$ is simply the length of the vector $Q^A$ in a Lorentzian lattice. Vanishing cosmological constant implies then that we are on the light-cone of this Lorentzian lattice. 
Realistically however, 
the configuration of the  $n^A$ 
should lead to a space-like $\Lambda$ since we have a positive cosmological constant, although very small. 
\begin{figure}[h!]
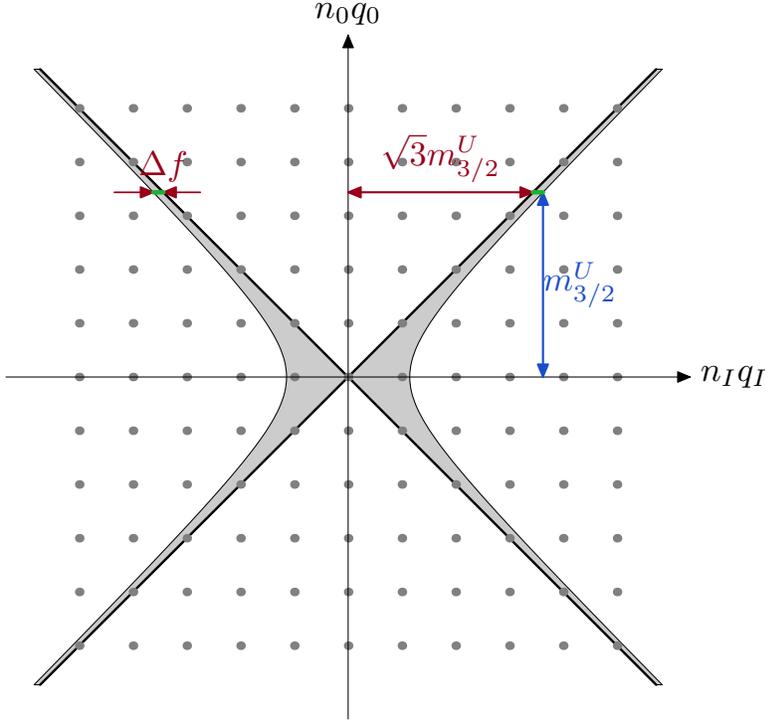
 
  \centering
  $\hbox{ \convertMPtoPDF{lattice2D.1}{1.3}{1.3} }$
    \caption{ The light cone of the Lorentzian lattice. Configurations residing inside the light-cone strip of width $\Delta f$ give rise to a universe with the observed value of the  cosmological constant. At the same time, these configurations should be at a distance $m_{3/2}^{U}$ from the origin, where $m_{3/2}^{U}$ is the gravitino mass of our universe.  } 
    \label{fig:awesome_image}
\end{figure}

Any space-like configuration $(n^0,n^I)$  in the above Lorentzian lattice that is on the light cone of width 
$\Delta \lambda$ leads to a universe with the observed value of the cosmological constant ($\Delta \lambda\sim 10^{-120}$ in Planck units). The condition that the cosmological constant possesses  a value close to the observed one is then written as 
\begin{equation}
0<\sum_{A=0}^N Q^A \eta_{AB} Q^B<\Delta \lambda. 
\end{equation}
The volume of the light-cone strip of width $\Delta f  = \Delta \lambda/(2 f)$  in the first quadrant\footnote{A transition to the region outside the given quadrant will be suppressed because it would imply an increase of the cosmological constant. This amounts to a $2^{-N}$ factor in \eqref{2N}.} at slant distance $f=\sqrt{3}m_{3/2}$ is then 
\begin{equation}
\label{2N}
V_{\rm lc}(m_{3/2})= 2^{-N-1} 3^{\frac{N-2}{2}}m_{3/2}^{N-2} \Omega_{N-1} \Delta \lambda,
\end{equation}
where $\Omega_{N-1}=2 \pi^{\frac{N}{2}}/\Gamma(N/2)$ is the volume of the unit $(N-1)$-dimensional sphere $S^{N-1}$.  The condition  to have a solution to the cosmological constant problem amounts therefore to having at least one point $(n^0,n^I)$ on the light cone. 
 There is an additional constraint though: we are interested in  those configurations inside the light-cone strip that have a large enough gravitino mass, in order to be consistent with the gravitino mass  $m_{3/2}^U$ in our universe. Therefore, in order to have a universe with a small cosmological constant and a gravitino mass $m_{3/2}^U$ we need
\begin{equation}
 D\prod_{A=0}^N q^A<V_{\rm lc}(m_{3/2}^U), \label{DA}
 \end{equation} 
 where the factor $D$ has been inserted to account for the large degeneracies expected \cite{BP}.
Hence, we get that the typical spacing of the scanning of the cosmological constant will be
\begin{equation}
 \Delta \lambda=\frac{2^{N+1} D\prod_{A=0}^N q^A}{3^{\frac{N-2}{2}}\Omega_{\!N\!-\!1}\, {m_{3/2}^U} ^{N-2}} .
\end{equation}
 Therefore, if $q^A\sim 10^{-a}, ~m_{3/2}^U\sim 10^{-b}$ in Planck  units, we find that the needed number of membranes to end up with a universe with the observed value of the cosmological constant is
\be
 N=\frac{120-a-2b}{a-b}\, . 
\ee
A typical value for $N$ is then $N \sim 100$ for $a,b \sim {\cal O}(1)$ (or ${\cal O}(10)$), as in \cite{BP}.

\section{Reheating}

The neutralization of the cosmological constant by membrane nucleation  leads to an empty universe, 
where all matter energy density has been diluted. 
A mechanism to overcome this issue is presented in \cite{BP}, 
and it requires the universe to enter a period of  inflation after the last nucleation. 
We show here how it can be reproduced in a manifest supersymmetric effective setup.

We have again a set of chiral three-form superfields $Y^I=(\Phi,Y^i)$ which satisfy the nilpotency conditions  
\be
\begin{split}
\Phi^2 &= 0, 
\\
\Phi \, Y^i &= 0, 
\end{split}
\ee
and by their coupling to the multiple membranes, as we have 
shown in the previous section, they lead to a supergravity realization of the BP mechanism.  
The inflaton will reside in the lowest component of the  chiral superfield ${\cal A }$ 
\be
\overline {\cal D}_{\dot \alpha}  {\cal A } = 0 . 
\ee 
Note that ${\cal A}$ is  a standard chiral superfield 
\be
{\cal A} = \phi + i b + \sqrt 2 \Theta \zeta + \Theta^2 F^A . 
\ee 
Here $\phi$ and $b$ are real scalars, 
$\zeta_\alpha$ is the physical fermion of the ${\cal A}$ multiplet 
and $F^A$ is the complex auxiliary field. 
To further simplify the model, 
we will use a constraint which 
has become very useful in supergravity cosmology \cite{Ferrara:2015tyn,Carrasco:2015iij,Dall'Agata:2015lek} 
for the description of the inflationary sector. 
The constraint we impose on ${\cal A}$ is 
\be
\Phi {\cal A} - \Phi \overline {\cal A } = 0.   
\ee
This constraint eliminates all the component fields of the chiral superfield ${\cal A}$, 
and leaves only the real scalar $a$ in the spectrum. 
Indeed, ignoring fermions, for the lowest component we have 
\be
{\cal A }|  = \phi \in \mathbb{R},  
\ee
while for the auxiliary field we have 
\be
-\frac14\left. {\cal D}^2 {\cal A }\right|  = 0. 
\ee
The function $\Omega$ we use now is 
\be
\label{omega2}
\Omega = -3 + \delta_{I  J}  Y^I \overline Y^J  -\frac{1}{4} \left( {\cal A} - \overline {\cal A}  \right)^2  
\ee
and the  superpotential is
\be
\label{WWW}
W =  \Phi \,  f( {\cal A} )   + g({\cal A}), 
\ee 
where the functions $f( {\cal A} )$ and $g({\cal A})$ are real, in the sense that $f^*(z) = f(z^*) $. 
These models have been studied in \cite{Ferrara:2015tyn,Carrasco:2015iij,Dall'Agata:2015lek} for standard chiral superfields.  
It is interesting that couplings  like \eqref{WWW} of the three-form superfield to the inflaton have been studied in \cite{Dudas:2014pva}. 
In this setup the multiple three forms are again  coupled to one membrane each, as in the previous section, 
and they give rise to the BP mechanism, 
and ${\cal A}$ will contain the inflationary sector. 
The bosonic sector related to ${\cal A}$ is 
\be
e^{-1} {\cal L}_{\cal A} = - \frac12 \, \partial^m \phi \partial_m \phi   
- \frac13 \left( M_1 \right)^2  + \delta_{IJ} F_1^I F_1^J  + 2 F_1 f(\phi)  
- 2 M_1 g(\phi) 
\ee 
and after we integrate out the auxiliary fields we find  
for the bosonic sector of the bulk theory 
\be
e^{-1} {\cal L} = -\frac12  R - \frac12 \, \partial^m \phi \partial_m \phi -  f^2(\phi) + 3 \, g^2(\phi) - \Lambda,   \label{inf}
\ee
where $\Lambda$ is given by \eqref{cosmo1}. 
Note that now also the function $g(\phi)$ will contribute to the gravitino mass \cite{Dall'Agata:2015lek}. 
For further discussion on supersymmetry breaking and nonlinear realizations in supergravity see 
\cite{Bergshoeff:2015tra,Hasegawa:2015bza,Dall'Agata:2014oka,Dall'Agata:2016yof,Antoniadis:2014oya,Ferrara:2014kva,
Ferrara:2015gta,Dudas:2015eha,Bandos:2015xnf}.

Let us  now see,  following \cite{BP}, how the  empty universe problem can be resolved in the present supersymmetric context. 
The scalar potential  is given by 
\be 
{\cal V}(\phi) = f^2(\phi) - 3 \, g^2(\phi), 
\ee  
where the functions  $f$ and $g$ are such that the slow-roll conditions are satisfied and ${\cal V}(0)=0$, $g(0)=0$. 
Assuming  large membrane tensions, 
the value of the cosmological constant before the penultimate step can be large enough to allow for the 
quantum fluctuations to dominate  over the  classical motion of the inflaton. If so, these fluctuations  kick the inflaton $\phi$ to random field values. 
As usual, since the quantum fluctuations of $\phi$ are typically given by 
\begin{equation}
|\delta \phi|\approx \frac{H(\phi)}{\sqrt{2}\, \pi}
\end{equation}
and its  classical change is 
\begin{equation}
|\Delta\phi|\approx \frac{{\cal V}'(\phi)}{2H(\phi)^2},
\end{equation}
the condition for the quantum fluctuations to dominate over the classical change $|\delta \phi|>|\Delta \phi|$ turns out to be (restoring $M_p$)
\begin{equation}
\Lambda_{\rm{eff}}^{3/2}>\left(\frac{M_p}{2}\right)^3{\cal V}'(\phi).  \label{empty}
\end{equation}
If $\Lambda_{\rm{eff}}$ is large just before the final membrane nucleation, the value of the inflaton $\phi$ will take a random value in the neighborhood of $\phi=0$. Then, after the final membrane 
 nucleation,  quantum fluctuations will not be important anymore and the classical dynamics will start dominating. The inflaton will then roll to its minimum at $\phi=0$, reheating the universe. 

To sum it up, before the penultimate step 
the field will take different values in the different regions of space, 
but only in some regions of space the inflaton will take the appropriate values to drive inflation 
for a sufficient number of e-folds, and subsequently reheat the universe. 
In these regions, 
once the final nucleation takes place, 
the classical motion of the inflaton will dominate over the quantum  fluctuation, and inflation will start. 
Our universe will emerge from one of these regions.  We do not have to add anything more than what was already discussed in \cite{BP}. We just want to stress that the condition for avoiding the empty universe problem is given in Eq. (\ref{empty}) whereas the rest of the details can be found in the original proposal in \cite{BP}.  
In addition, 
after the final nucleation takes place 
the life-time of the state with configuration $n^A$ can be easily made sufficiently larger than the life-time of our universe \cite{BP}.

\section{Conclusions}

In this paper we have studied   the scanning of the supersymmetry breaking scale and the gravitino mass induced by membrane nucleation, which  may give rise to the dynamical neutralization of the cosmological constant. 
The mechanism relies on the fact that, by using the three-form minimal supergravity multiplet, 
 the cosmological constant  becomes a dynamical variable since  
both the gravitino mass and  the supersymmetry breaking scale make jumps.

We have explicitly constructed a realization of this  supergravity   setup by showing that 
the supersymmetry braking scale and the gravitino mass can be scanned by membrane nucleation which alters their values. 
The favored processes are the ones which lead to a lower cosmological constant, and therefore 
the supersymmetry breaking scale will principally tend to decrease while the gravitino mass will tend to increase.  
This process   stops as soon as we end up with a small cosmological constant, with value close to the observed one. At that point, the supersymmetry breaking scale can be large and proportional to the gravitino mass such that the cosmological constant is almost zero 
and further membrane nucleation is very suppressed. 
Our findings show that, even though the  minimal 4D ${\cal N}=1$ supergravity is only the low energy limit of string theory (or M-theory), 
it is still a powerful tool and can be used to study non-perturbative aspects of the dynamics.

\section*{Acknowledgements} 
We thank G.\ Dall'Agata, L.\ Martucci and R.\ von Unge for discussion. 
F.F. is supported in part by the MIUR grant RBFR10QS5J (STaFI) and by the Padova University Project CPDA119349.  A.R. is supported by the Swiss National Science Foundation (SNSF), project ``Investigating the Nature of Dark Matter" (project number: 200020-159223).

\end{document}